\documentclass[12pt]{article}
\usepackage{amsmath,amssymb,bm,epsfig,afterpage,tabularx}
\usepackage{cite}
\usepackage{here}
\usepackage{color}

\renewcommand{\thefootnote}{\fnsymbol{footnote}}

\begin{document}
\title{
\begin{flushright}
\begin{minipage}{0.2\linewidth}
\normalsize
WU-HEP-17-06 \\*[50pt]
\end{minipage}
\end{flushright}
{\Large \bf 
Study of dark matter physics  \\
in non-universal gaugino mass scenario}
\\*[20pt]}
\author{
Junichiro~Kawamura$^a$\footnote{
E-mail address: junichiro-k@ruri.waseda.jp} \ and
Yuji~Omura$^b$\footnote{
E-mail address: yujiomur@kmi.nagoya-u.ac.jp}\\*[20pt]
{\it \normalsize 
$^a$Department of Physics, Waseda University, 
Tokyo 169-8555, Japan} \\
{\it \normalsize 
$^b$Kobayashi-Maskawa Institute for the Origin of Particles and the Universe (KMI), } \\
{\it \normalsize
Nagoya University, Nagoya 464-8602, Japan} \\*[50pt]
}
\date{
\centerline{\small \bf Abstract}
\begin{minipage}{0.9\linewidth}
\medskip 
\medskip 
\small
 We study dark matter physics in the Minimal Supersymmetric Standard Model
 with non-universal gaugino masses at the unification scale.
 In this scenario, the specific ratio of wino and gluino masses
 realizes the electro-weak scale naturally and achieves 125 GeV Higgs boson mass. Then, relatively light higgsino is predicted
 and the lightest neutral particle, that is dominantly given by the neutral component of higgsino, is a good dark matter candidate. The direct detection of the dark matter
 is sensitive to not only a higgsino mass
 but also gaugino masses significantly. 
 The upcoming XENON1T experiment excludes the parameter region where bino  or gluino is lighter than about 2.5 TeV
 if the higgsino and the gaugino mass parameters have same signs.  
 We see that the direct detection of dark matter 
 gives stronger bound than the direct search at the
 LHC experiment when higgsino sizably contributes to the 
 dark matter abundance. 
\end{minipage}
}

\begin{titlepage}
\maketitle
\thispagestyle{empty}
\clearpage
\tableofcontents
\thispagestyle{empty}
\end{titlepage}

\renewcommand{\thefootnote}{\arabic{footnote}}
\setcounter{footnote}{0}

\section{Introduction}
Supersymmetry (SUSY) is a promising candidate for physics beyond the Standard Model (SM).
The supersymmetric extension predicts the superpartners of the SM particles,
and the masses of the SUSY particles are expected to be at least TeV-scale,
in order to explain the origin of the electroweak (EW) scale.\footnote{See for reviews e.g.~\cite{Martin:1997ns,Chung:2003fi}.} In the Minimal Supersymmetric Standard Model (MSSM),
there is a supersymmetric mass parameter, what is called $\mu$-parameter, for higgsino that is the superpartner of Higgs bosons. In order to realize the EW scale without fine-tuning, $\mu$-parameter should be EW-scale.
Besides, the lightest particle in the MSSM becomes stable because of R-parity,
so that higgsino becomes a good dark matter (DM) candidate if there is no lighter SUSY particle. 
So far, a lot of efforts are devoted to the SUSY search in the collider experiments and the
dark matter observations~\cite{Jungman:1995df}. 
There are no decisive signals of the SUSY particles, but higgsino is still one of the possible and attractive 
DM candidates that reveal the origin of the EW scale.

In the MSSM, there are a lot of parameters, so that we can consider many possibilities of the mass spectrum 
for the SUSY particles. The direct searches for the SUSY particles as well as the 125~GeV Higgs boson mass measurement
at the LHC~\cite{Aad:2015zhl},
however, constrain the parameter space strictly. 
It is getting very difficult to construct SUSY models, as long as the explanation of the EW scale is not discarded.
One possible setup to achieve both the 125 GeV Higgs boson mass and the explanation of the EW scale
is known as the Non-Universal Gaugino Masses (NUGM)
scenario~\cite{Abe:2007kf,Abe:2012xm}. In this scenario, 
a suitable ratio of the wino mass to the gluino mass achieves the EW scale
and the 125~GeV Higgs boson mass. 
Then, the $\mu$-parameter is predicted to be close to the EW scale.
The current status and the future prospect of the discovery of the SUSY particles at the LHC
have been investigated in this scenario~\cite{Abe:2015xva,Kawamura:2016drh,JK_PhDthesis}. 
We find that the superpartners of top quark and gluon, what are called top squark and gluino, are promising particles to test this scenario. 
Expected reaches of these SUSY particles decaying to higgsinos 
are studied in Refs.~\cite{Baer:2016bwh,Baer:2016wkz}.

Note that there are some models that lead such a ratio of the gauginos.
One possibility is
the mirage mediation~\cite{Choi:2004sx,Choi:2005ge,Choi:2005uz},
that is a mixture of the moduli mediation~\cite{Brignole:1995fb,Brignole:1997dp}
and anomaly mediation~\cite{Randall:1998uk,Giudice:1998xp}.
The phenomenology of the mirage mediation
is discussed before the Higgs boson discovery
in Refs.~\cite{Choi:2006xb,Kitano:2005wc,Choi:2006im,Cho:2007fg,Nagai:2007ud,Nakamura:2008ey,Choi:2009jn,Holmes:2009mx,Altunkaynak:2010xe,Everett:2008ey,Everett:2008qy}
and after that in Refs.~\cite{Asano:2012sv,Kobayashi:2012ee,Abe:2014kla,Hagimoto:2015tua,Everett:2015dqa,Barger:2015xqk,Baer:2016hfa}.  
There are some works to realize the ratio of the gauginos
in the GUT models~\cite{Younkin:2012ui}  
and superstring models~\cite{Blumenhagen:2006ci}.

In this kind of SUSY models, higgsino is light because of the explanation of the origin of the EW scale,
and the SUSY particle is expected to be discovered in experiments.
There are neutral and charged components in higgsino, and the neutral component 
mixes with bino and wino, and the charged component mixes with wino.\footnote{Wino and bino
are the superpartners of $SU(2)_L$ and $U(1)_Y$ gauge bosons, respectively. }
In our scenario, the gauginos are relatively heavy, so that 
all components of higgsino are light and almost degenerate; in fact,
the mass difference is a few GeV~\cite{Abe:2015xva,Kawamura:2016drh}.
Then, higgsino is hard to be detected at the LHC due to the certainly small mass differences.
On the other hand,
dark matter direct detection experiments can efficiently observe higgsinos,
if the neutral component of higgsino slightly mixes with the gauginos
and dominates over our universe. 
It is also interesting that the higgsino mass should be lighter than
about 1 TeV, if higgsino is thermally produced. 
Then, our DM mass, that mainly comes from the neutral component of higgsino,
is predicted to be between the EW scale and 1 TeV.

In this paper,
we study dark matter physics in the NUGM scenario.
Direct detection experiments are sensitive to not only
the higgsino mass itself, but also the gaugino masses,   
because the higgsino-gaugino mixing
gives the most significant contribution to the detection rate. 
We also discuss the constraints from the LHC experiments,
based on the results in Refs.~\cite{Abe:2015xva,Kawamura:2016drh,JK_PhDthesis}.
We explicitly show the exclusion limit and the future prospect on the plane of the higgsino and the gaugino masses. In the end, we find that this scenario can be fully covered by the future experiments, as far as  the gluino mass is below 2.5 TeV in a certain parameter set.

This paper is organized as follows.
The NUGM scenario is reviewed in Section 2, and
we discuss dark matter physics in Section 3.
The results of numerical calculations are shown in Section 4.
Section 5 is devoted to conclusion.

\section{NUGM scenario}
\subsection{Review of NUGM} 
The NUGM scenario is known as one of the attractive SUSY models to realize
$\mu$-parameter near the EW scale 
and the 125~GeV Higgs boson mass simultaneously.
The $\mu$-parameter is related to the EW symmetry breaking scale
through the minimization condition for the Higgs potential as
\begin{eqnarray}
 \label{eq-EWSB}
 m_Z^2 \simeq 2 |m_{H_u}^2| - 2|\mu|^2,
\end{eqnarray}
where $m_Z$ is the Z-boson mass
and $m_{H_u}^2$ is the soft scalar mass squared for the up-type Higgs boson. 
This relation shows that $|\mu|^2$ and $ |m_{H_u}^2|$ should be around the EW scale 
to avoid the fine-tuning between those parameters. 
The $\mu$-parameter is an unique SUSY-preserving parameter in the MSSM.
On the other hand, all other dimensional parameters softly 
break SUSY and would be originated
from some mediation mechanisms of SUSY breaking: i.e., the soft SUSY breaking terms would have same origin. 
Let us assume that the all ratios of soft SUSY breaking parameters are fixed 
by some mediation mechanisms and the overall scale is given by $M_0$. 
In this assumption, Eq. (\ref{eq-EWSB}) corresponds to the
relation between $\mu$ and $M_0$.
In Ref.~\cite{Barbieri:1987fn}, the parameter,  $\Delta_x$, to measure the sensitivity of the parameter $x$ to the EW scale is introduced:
\begin{eqnarray}
 \Delta_x = \left|\frac{\partial \ln{m_Z^2}}{\partial \ln{x^2}}\right|\ (x = \mu, M_0).
\end{eqnarray}
Since $m_{H_u}^2(m_{\rm SUSY})$ is expressed as a quadratic polynomial function
of the boundary conditions, 
we can derive $\Delta_\mu + \Delta_{M_0}=1$ at the tree-level 
and $\Delta_{\mu} \simeq \Delta_{M_0}$ is satisfied.
Thus the tuning of the $\mu$-parameter represents 
the degree of tuning to realize the EW symmetry breaking in the model.
From the relation Eq.~(\ref{eq-EWSB}),
the tuning measure of the $\mu$-parameter can be written as
$\Delta_\mu = 2|\mu|^2/m_Z^2$ up to radiative corrections to the condition, 
so that 
small $|\mu|$ is simply required to avoid the fine-tuning in this assumption.
The details of this kind of discussions in the NUGM scenario are shown
in Refs.~\cite{JK_PhDthesis,preparation}.  
We proceed to study collider and dark matter phenomenology
with the NUGM in this assumption.

In this paper,
we assume universal soft scalar mass $m_0$ and A-term $A_0$, 
while the gaugino masses $M_{1,2,3}$ are non-universal
at the gauge coupling unification scale ($\simeq 10^{16}$ GeV).
We assume the ratio of two Higgs vacuum expectation values (VEVs)  
$\tan\beta \equiv \langle H_u\rangle/\langle H_d\rangle =10$ throughout this paper.  
The soft mass squared $m_{H_u}^2$ at $m_{\rm SUSY}=1$ TeV
relates to the boundary conditions at the unification scale as
\begin{eqnarray}
 m_{H_u}^2(m_{\rm SUSY})&\simeq&0.005M_1^2 - 0.005 M_1M_2 + 0.201 M_2^2
                      -0.021 M_1M_3 - 0.135 M_2M_3  \\ \nonumber
                   && - 1.57 M_3^2+ A_0 (0.011 M_1 + 0.065 M_2 + 0.243 M_3 - 0.099 A_0)
	              - 0.075 m_0^2. 
\end{eqnarray}
This relation shows that
the contribution from the gluino mass is dominant among the renormalization group (RG) effects,
but we find that the gluino mass contribution can be canceled by the RG effects from the other gaugino masses $M_{1,2}$. 
In particular, the $M_2^2$ term cancels the $M^2_3$ term if the ratio of $M_2/M_3$ satisfies $M_2/M_3\simeq$ 3-4. 
Similarly,
the top squark mass parameters $m^2_{\tilde{t}_L}$, $m^2_{\tilde{t}_R}$ and $A_t$
at $m_{\rm SUSY}=1$ TeV
are related to the boundary conditions as 
\begin{eqnarray}
 m_{\tilde{t}_L}^2(m_{\rm SUSY})
  &\simeq& -0.007 M_1^2   -0.002 M_1M_2 + 0.354 M_2^2
                    -0.007 M_1 M_3 -0.051 M_2 M_3 + 3.25 M_3^2 \nonumber \\ \label{eq-mQ}
        && +(0.004 M_1+0.025 M_2+0.094 M_3-0.039 A_0) A_0+ 0.622 m_0^2, \\
 m_{\tilde{t}_R}^2(m_{\rm SUSY})
  &\simeq&  0.044 M_1^2 -0.003 M_1 M_2 - 0.158 M_2^2
                    -0.014 M_1 M_3 -0.090 M_2 M_3 +2.76 M_3^2 \nonumber \\ \label{eq-mu}
        && +(0.008 M_1+0.044 M_2+0.162 M_3-0.066 A_0) A_0+ 0.283 m_0^2, \\
A_t(m_{\rm SUSY})   &\simeq&  -0.032 M_1 - 0.237 M_2 - 1.42 M_3 + 0.277 A_0.  	     
\label{eq-At}
\end{eqnarray}
We see that
$A_t(m_{\rm SUSY})$ increases and $m_{\tilde{t}_R}^2(m_{\rm SUSY})$ decreases as the wino mass $M_2$ increases. 
Note that the latter effect is induced by the top Yukawa coupling.
As a result,
the ratio $A_t^2/\sqrt{m_{\tilde{t}_L}^2m_{\tilde{t}_R}^2}$ increases and
the SM-like Higgs boson mass around 125 GeV can be achieved due to the relatively large wino.

\subsection{Mass spectrum of NUGM}
We see that the suitable wino-to-gluino mass ratio reduces the $\mu$-parameter
and also enhances the Higgs boson mass.
Besides, some of sparticle masses are within reaches of the LHC experiment
thanks to the sizable left-right mixing of the top squarks \cite{Abe:2015xva,Kawamura:2016drh}.

When the wino mass is large,
left-handed sparticles become heavy due to the RG evolution. 
The right-handed slepton masses are determined by the bino mass, 
while the right-handed squark masses mainly depend on both the gluino and bino masses.
The bino mass plays a crucial role in shifting the top squark mass, as well.  
This means that the bino mass have to be so heavy that the top squark mass is enough heavy
to be consistent with the LHC results.

Another important point derived from the relatively heavy bino and wino
is that the mass differences among the components of higgsino become small. 
The mass differences are induced by the mixing with higgsino and gauginos,
so that these are suppressed by the bino and wino masses
as explicitly shown in next section.
The mass differences among the components of higgsino
are typically 2 GeV as shown in Ref.~\cite{Abe:2015xva}. 
This small mass difference makes it difficult to detect higgsino directly at the LHC, 
because their daughter particles are too soft to be distinguished from backgrounds
and their lifetimes are too short to be recognized as charged tracks
unlike the case that wino is the lightest SUSY particle (LSP)~\cite{Ibe:2006de}~\footnote{There are recent works to study searching for charged higgsinos that exploit their relatively long lifetime~\cite{Mahbubani:2017gjh,Fukuda:2017jmk}.
}.  
This feature also indicates that we can treat all of the particles from higgsino as invisible particles at the LHC.

Let us summarize the important features of our mass spectrum discussed below:
\begin{itemize}
\item All gauginos are ${\cal O}(1)$~TeV.
\item The higgsino mass is between the EW scale and 1~TeV,
       and the mass differences are ${\cal O}(1)$~GeV. 
\item Right-handed top squark is relatively light. 
\end{itemize}

\subsection{LHC bounds}
In our scenario, the top squark and the gluino are the good candidates to be detected at the LHC. 
The current exclusion limit and the future prospect have been studied in Refs.~\cite{Abe:2015xva,Kawamura:2016drh,JK_PhDthesis}.

In the NUGM scenario,
a top squark decays as
$\tilde{t}_1\rightarrow t\tilde{\chi}_{1,2}^0/b\tilde{\chi}_1^\pm$
where each branching fraction is $50\%$ as long as
the mass difference between the top squark and each of the higgsino-like particles
is significantly larger than the top quark mass. 
Note that the neutralinos consist of higgsino that slightly mixes with wino and bino in our scenario.   
The relevant top squark searches at the LHC are discussed in
Ref.~\cite{LHC13:sb} and Ref.~\cite{LHC1313_hadMET}.  
The former analysis aims to a pair of bottom squarks that decay as 
$\tilde{b}_1\tilde{b}_1\rightarrow b\tilde{\chi}^0 b\tilde{\chi}^0$.
This gives same signal as 
$\tilde{t}_1\tilde{t}_1\rightarrow b\tilde{\chi}^\pm b\tilde{\chi}^\pm$
in the NUGM scenario. 
The latter analysis aims to
hadronically decaying top squarks,
$\tilde{t}_1\tilde{t}_1\rightarrow t \tilde{\chi}^0 t \tilde{\chi}^0$
$ \rightarrow bjj\tilde{\chi}^0 bjj\tilde{\chi}^0$. 
In Ref. \cite{LHC1313_hadMET},
the signal regions require more than 4 jets,
where 2 of these should be b-tagged. 
Such signal regions will be sensitive to events 
$\tilde{t}_1 \tilde{t}_1 \rightarrow t (\rightarrow bjj)\ \tilde{\chi}^0 b \tilde{\chi}^\pm$
in the NUGM scenario,
although this analysis is not completely optimized.    
This decay pattern is realized
in almost half of the events with the pair produced top squarks 
if the mass difference between the top squark and higgsino is enough large. 
Thus this channel that targets to the hadronically decaying top squark
is sensitive to the large mass difference region,
while the former channel that targets to
bottom squarks decaying to a bottom quark and a neutralino
is sensitive to the mass degenerate region. 
Referring the analysis in Ref.~\cite{JK_PhDthesis}, top squark lighter than 800 GeV is excluded if $\mu\lesssim200$ GeV is satisfied, and top squark lighter than 600 GeV is excluded in the range with $200$ GeV $\lesssim \mu \lesssim 270$ GeV. There is no exclusion limit for top squarks if $\mu$ is greater than 270 GeV.

In present scenario,
a gluino decays as
$\tilde{g}\rightarrow t\tilde{t}_1\rightarrow t+t\tilde{\chi}^0/b\tilde{\chi}^\pm$. 
Hence, the the signal from the gluino pair production
is expected to have 4 b-tagged jets, jets/leptons coming from 2-4 W-bosons and large missing energies
in the final state.
The analysis in Ref.~\cite{LHC13:glu} aims to this type of signals,
and we refer the exclusion limit obtained in Ref.~\cite{JK_PhDthesis}. 
Gluino lighter than 1.8 TeV is excluded if the $\mu$-parameter is less than 800 GeV.
The bound is relaxed if the mass difference is smaller than about 300 GeV.

Note that there is another channel, $\tilde{g} \rightarrow g\tilde{\chi}^0$,
that is induced by the top squark loop. 
If the mass difference
between gluino and higgsino is near or less than the top quark mass, this decay channel becomes important.
We need to consider the limits based on data such as Ref.~\cite{LHC:2jMET}, but it is beyond the scope of this paper.

Let us comment on the case with light bino.
If gluino is enough heavy, bino can be as light as higgsino and top squark can also decay to bino. 
The decay is, however, usually suppressed unless bino is significantly lighter than higgsino
because the coupling of bino with top squark is much weaker
than the one of higgsinos because of the top Yukawa coupling.
Such a light bino is less attractive from the experimental point of view.  
If the bino mass is light, 
gluino has to be much heavier than
the experimental reach in order to shift the top squark mass. 
Then, the light bino case would be unfavorable from 
the naturalness point of view. 
Furthermore,
it is known that bino LSP tends to overclose the universe 
and some dilution mechanisms are necessary.

\section{Dark matter physics}
\subsection{Neutralino sector}
In our study, we assume that the signs of all the gaugino masses are positive and the sign of the $\mu$-parameter is either negative or positive. 
After the EW symmetry breaking, gauginos and higgsino are mixed each other.
The neutralino mass matrix in a basis of
$\psi =(\tilde{B},\tilde{W},\tilde{H}_d^0,\tilde{H}_u^0)$ is given by
\begin{eqnarray}
 M_{\tilde{\chi}} =
  \begin{pmatrix}
   M_1 & 0   & -c_\beta s_W m_Z & s_\beta s_W m_Z  \\
   0   & M_2 & c_\beta c_W m_Z  & -s_\beta c_W m_Z \\
   -c_\beta s_W m_Z & c_\beta c_W m_Z & 0 & -\mu    \\
   s_\beta s_W m_Z  & -s_\beta c_W m_Z & -\mu & 0  
  \end{pmatrix}, 
\end{eqnarray}
where $c_\beta=\cos\beta$, $s_\beta=\sin\beta$, $c_W=\sin\theta_W$
and $s_W = \sin\theta_W$ are defined and 
$\theta_W$ is the Weinberg angle. 
This matrix is diagonalized by an unitary matrix $N$ as
\begin{equation}
 \psi_i = N_{ij} \tilde{\chi}_j ~~{\rm and}~~
 N^\dagger M_{\tilde{\chi}} N =
  {\rm diag} (m_{\tilde{\chi}_1},m_{\tilde{\chi}_2},m_{\tilde{\chi}_3},m_{\tilde{\chi}_4}).  
\end{equation}
The masses, $m_{\tilde{\chi}_1}$, $m_{\tilde{\chi}_2}$, $m_{\tilde{\chi}_3}$ and $m_{\tilde{\chi}_4}$ approach to $M_1$, $M_2$, $\mu$, and $-\mu$ in the limit that $m_Z$ is vanishing, respectively.
The mass eigenstate $\tilde{\chi}_3$ ($\tilde{\chi}_4$) becomes the lightest one 
if the $\mu$-parameter is positive (negative) and $|\mu| < M_1, M_2$.

The neutralino-neutralino-Higgs coupling, 
$\mathcal{L} \ni (1/2) \lambda_{hnn} h \overline{\tilde{\chi}}_n \tilde{\chi}_n$, 
is given by
\begin{eqnarray}
 \label{eq-LLH}
 \lambda_{hnn} = g (s_\alpha N_{3n}+c_\alpha N_{4n})(N_{2n}-t_W N_{1n}),  
\end{eqnarray}
where $t_W$, $s_\alpha$ and $c_\alpha$ are short for $\tan\theta_W$, $\sin\alpha$ and $ \cos\alpha$, respectively.
 $\alpha$ is a mixing angle of the Higgs boson. 
The mixing matrix is given by 
\begin{eqnarray}
 \label{eq-N1}
  (N_{11},N_{21},N_{31},N_{41})
  &=& \left(1, \, 0, \, -\frac{m_Zs_W(c_\beta M_1+s_\beta\mu)}{M_1^2-\mu^2}
              ,\,   \frac{m_Zs_W(c_\beta \mu+s_\beta M_1)}{M_1^2-\mu^2}\right), \\
 \label{eq-N2}
  (N_{12},N_{22},N_{32},N_{42})
  &=& \left(0,\, 1,\, \frac{m_Zc_W(c_\beta M_2+s_\beta\mu)}{M_2^2-\mu^2},\,
               -\frac{m_Zc_W(c_\beta \mu+s_\beta M_2)}{M_2^2-\mu^2}\right), \\
 \label{eq-N3}
 (N_{13},N_{23},N_{33},N_{43})
 &=& \frac{1}{\sqrt{2}}\left( \frac{m_Zs_W(c_\beta+s_\beta)}{M_1-\mu}, \,
                                - \frac{m_Zc_W(c_\beta+s_\beta)}{M_2-\mu}, \,1, \, -1\right), \\ 
 \label{eq-N4}
 (N_{14},N_{24},N_{34},N_{44})
 &=& \frac{1}{\sqrt{2}}\left( \frac{m_Zs_W(c_\beta-s_\beta)}{M_1+\mu}, \,
                                - \frac{m_Zc_W(c_\beta-s_\beta)}{M_2+\mu}, \,1,\, 1 \right), 
\end{eqnarray}
where $m_Z \ll |M_{1,2}\pm\mu|$ is assumed.

\subsection{Thermal relic abundance}
It is known that
the thermal relic density of the purely higgsino LSP saturates the universe
when the higgsino mass is about 1 TeV~\cite{Cirelli:2005uq,Cirelli:2007xd}. 
If we assume that there is no dilution effect
after the thermal production of the LSP,
the higgsino-like LSP heavier than 1 TeV
overcloses the universe
and is cosmologically excluded
unless the higgsino and another sparticle, such as a top squark,
are so degenerate that co-annihilation processes between them reduce the relic density.

Let us comment on possibilities
that gauginos contribute to dark matter considerably. 
In our scenario, the wino mass should be as large as the gluino mass at the TeV scale
and it hardly contributes to the dark matter. 
The bino mass can be as light as the higgsino mass
if the gluino mass is enough large to keep the top squark mass. 
It was interesting that the well-tempered bino-higgsino LSP
explains the observed abundance in the thermal scenario~\cite{ArkaniHamed:2006mb},  
but most of parameter space has been already excluded
by the direct detections as will be discussed later~\footnote{
There are narrow regions where the thermal bino-higgsino LSP explains the abundance
by the Higgs- or Z-boson resonances without tension with the DM direct detection experiments~\cite{Hamaguchi:2015rxa}.}.

In our scenario, the relic DM abundance thermally produced 
may not be sufficient to satisfy the observed DM abundance in our universe.
When we denote the relic abundance of the LSP as
$\Omega_\chi h^2$, we can simply consider two possibilities to saturate the observed value, $\Omega_{\rm obs} h^2= 0.1188\pm0.0001$ \cite{Planck}:
\begin{enumerate}
\renewcommand{\labelenumi}{(\Alph{enumi})}
\item $\Omega_\chi h^2$ is only given by the thermal production, and $\Omega_\chi h^2 \leq \Omega_{\rm obs} h^2$ is satisfied.
\item $\Omega_\chi h^2 = \Omega_{\rm obs} h^2$ is always satisfied, assuming non-thermal production of LSP works.
\end{enumerate} 

In the case (A), what is called $thermal$ $scenario$, the LSP may not saturate our universe, depending on the 
parameter region. Then, we need other dark matter candidates such as axion to achieve the observed relic abundance of DM. 

In the case (B), what is called $non$-$thermal$ $scenario$, we simply assume that the LSP 
dominates our universe and satisfies $\Omega_\chi h^2 = \Omega_{\rm obs} h^2$.
We do not explicitly calculate the relic abundance, but several mechanisms for
the non-thermal productions have been proposed so far.
For instance, it is known that
the decays of long-lived heavy particles, such as gravitino, saxion and moduli field,
can significantly produce the LSP after the LSP is frozen out from the thermal bath
~\cite{Kohri:2005ru,Baer:2011uz,Baer:2014eja,Allahverdi:2013noa}.

Note that the important difference of the two scenarios is whether $\Omega_\chi h^2 < \Omega_{\rm obs} h^2$
is allowed or not. In our study, we estimate the thermal relic density of the LSP,
and we exclude the region with $\Omega_\chi h^2 > \Omega_{\rm obs} h^2$~\footnote{
Note that this region is not truly excluded in the non-thermal scenario,
but such region satisfies $|\mu|\gtrsim 1.0$ TeV that is less attractive
from both the testability and the naturalness point of view.}.  
When we estimate the direct detection rate of DM, the abundance of the LSP is important.
Then we draw the exclusion limits of both cases.

\subsection{direct detection}
The direct detection for dark matter
is a promising way to probe the neutralino sector of the MSSM. 
The current limits on the spin-independent and spin-dependent cross sections
are given by the XENON100~\cite{Garny:2012it,Aprile:2012nq,Aprile:2013doa}, LUX \cite{LUX2015,LUX2016},
PANDAX-II \cite{Panda,Fu:2016ega} and PICO~\cite{Amole:2015pla,Amole:2016pye}.
The XENON1T \cite{XENON1T} and LZ \cite{Akerib:2015cja} will cover wider range in near future.

Let us discuss spin-independent cross section of neutralino scattering with nucleons. 
Note that the limits on the gaugino masses 
from the spin-independent cross section
are stronger than those  from the spin-dependent cross section in most cases.

At tree-level,
spin-independent scatterings are induced   
by the t-channel Higgs boson exchange
and the s-channel squark exchange. 
Since only one top squark is light in the NUGM scenario, 
the latter contribution is negligibly small. 
The mixing between gauginos and higgsino are important 
in the Higgs boson exchange,
because the LSP-LSP-Higgs coupling in the mass eigenstate basis 
is originated from the gaugino-higgsino-Higgs couplings in the gauge eigenstate basis.
In the limit of $m_Z \ll |M_{1,2}\pm \mu|$,  
the mixing effects are suppressed by $m_Z/|M_{1,2}\pm\mu|$
as shown in Eqs. (\ref{eq-N3}) and (\ref{eq-N4}).

It has been shown that there is a parameter set to lead vanishing gaugino-higgsino mixing,
what is called the blind spot \cite{Cheung:2012qy}.
As we see Eqs.(\ref{eq-LLH}), (\ref{eq-N1}) and (\ref{eq-N2}),
the mixing is proportional to $M_{1,2} + \mu \sin 2\beta$, 
so that the mixing vanishes when the relative signs of $M_{1,2}$ and $\mu$
are opposite, and $|M_{1,2}| \lesssim |\mu|$ and $\tan\beta \gtrsim 1$ are satisfied.
Thus the blind spot appears only in the gaugino-like LSP scenario.

Note that the mixing is suppressed when the LSP is higgsino-like
and signs of $\mu$ and $M_{1,2}$ are opposite,
as we can see from Eqs.(\ref{eq-N3}) and (\ref{eq-N4}). 
Since the mixing is proportional to $1\pm\sin2\beta$,
smaller $\tan\beta$ induces larger enhancement (suppression) for the same (opposite) sign.
We need $\tan\beta \gtrsim 10$ in order to realize the SM-like Higgs boson mass
unless the sparticle masses are much heavier than 1 TeV,
so that such effect is at most $~20\%$-level.  
Thus we conclude that the gaugino-higgsino mixing is sizable and
the factor, $1\pm\sin2\beta$, leads significant difference between the positive and the negative $\mu$-parameter cases
in the DM scattering cross section.

The spin-independent cross section per nucleon at the tree-level can be written as
\begin{eqnarray}
 \sigma_N^{\rm SI} = \frac{g^2}{4\pi} \frac{m_N^4}{m_h^4m_W^2}
                     \left(1+\frac{m_N}{m_\chi} \right)^{-2}
                     \left[\frac{2}{9}+\frac{7}{9}\sum_{q=u,d,s}f^N_{T_q} \right]^2
		     \lambda_{h\chi\chi}^2,  
 \label{eq-SIsimple}
\end{eqnarray}
where $m_N$ is the nucleon mass
and $m_N f_{T_q}^N = \langle N|m_q \bar{q}q |N\rangle$.
In the decoupling limit $m_A \gg m_Z$ that is a good approximation for our case, 
using Eqs. (\ref{eq-N3}) and (\ref{eq-N4}), 
the LSP-LSP-Higgs coupling $\lambda_{h\chi\chi}$ is derived from Eq. (\ref{eq-LLH}):
\begin{eqnarray}
 \label{eq-chxx}
 \lambda_{h\chi\chi} = \frac{g}{2}(1\pm s_{2\beta}) c_W
  \left(\frac{m_Z}{M_2-|\mu|} + t_W^2 \frac{m_Z}{M_1-|\mu|}\right), 
\end{eqnarray}
where $\pm$ corresponds to a sign of the $\mu$-parameter.

\newcommand{\none}{\tilde{\chi}^0_1}
 \newcommand{\ntwo}{\tilde{\chi}^0_2}
 \newcommand{\nthr}{\tilde{\chi}^0_3}
 \newcommand{\nfou}{\tilde{\chi}^0_4}
 \newcommand{\xone}{\tilde{\chi}^\pm_1}
 \newcommand{\xtwo}{\tilde{\chi}^\pm_2}
 \newcommand{\ethr}{\times 10^{-3}}
 \newcommand{\eone}{\times 10^{-1}}
 \newcommand{\sv}{\langle\sigma v\rangle_0 \times10^{25} [\rm cm^3/s]}
 \newcommand{\BRw}{{\rm Br}(\chi\chi \rightarrow W^+W^-)}
 \newcommand{\BRz}{{\rm Br}(\chi\chi \rightarrow ZZ)}
 \newcommand{\SD}{\sigma_{\rm SD}\times10^{-6 }   [{\rm pb}]}
 \newcommand{\SI}{\sigma_{\rm SI}\times10^{-11}   [{\rm pb}]}
 \newcommand{\SIh}{\sigma^h_{\rm SI}\times10^{-11}[{\rm pb}]}

 \begin{table}[!t]
  \centering
  \caption{Values of boundary conditions at the unification scale $M_U$,
  Higgs boson masses, sparticle masses and 
  dark matter observables at several sample points.}
  \vspace{0.5cm} 
  \begin{tabular}{|c|cc|cc|} \hline
   input      [GeV]& (a)  &  (b) &  (c)  & (d)  \\ \hline 
   $\mu$           & -250 & 250  & -1000 & 1000 \\     
   $M_1(M_U)$      & 10000& 10000&   5000& 5000 \\     
   $M_3(M_U)$      & 1000 & 1000 &   1500& 1500 \\
   $m_0(M_U)$      & 1000 & 1000 &  1000 & 1000 \\ \hline  
   output [GeV]    &      &      &       &      \\ \hline  
   $M_2(M_U)$      &4223  & 4175 &  4698 & 4504 \\  
   $A_0(M_U)$      &-2378 & -2325& -1916 & -1657\\ \hline
   mass [GeV]      &      &      &       &      \\ \hline
   $m_h$           & 125.0& 125.0& 125.0 & 125.0\\ 
   $m_A$           &  3349& 3326 & 3351  & 3248 \\ 
  $m_{\tilde{t}_1}$&  1606& 1636 & 1431  & 1581 \\ 
  $m_{\tilde{t}_2}$&  2780& 2762 & 3582  & 3520 \\ 
  $m_{\tilde{g}}$  &  2250& 2250 & 3225  & 3223 \\ 
      $m_{\none}$  & 258.8& 255.7& 1016  & 1013 \\ 
      $m_{\ntwo}$  & 260.5& 258.3& 1019  & 1017 \\ 
      $m_{\nthr}$  & 3438 & 3400 & 2239  & 2237 \\ 
      $m_{\nfou}$  & 4455 & 4454 & 3839  & 3682 \\ 
      $m_{\xone}$  & 260.5& 257.1& 1018  & 1015 \\ 
      $m_{\xtwo}$  & 3439 & 3400 & 3840  & 3682 \\ \hline 
  observables      &      &      &       &      \\ \hline
  $\Omega_\chi h^2$&7.82$\ethr$&7.58$\ethr$
	                         &1.14$\eone$&1.16 $\eone$ \\
   $\sv$           & 1.39 & 1.42 & 0.104 & 0.105 \\
   $\BRw$          & 0.533& 0.535& 0.488 & 0.489 \\
   $\BRz$          & 0.436& 0.435& 0.408 & 0.407 \\ \hline
  $\SD$            &1.096 & 1.138& 0.1677&0.1757 \\
  $\SI$            &3.499 & 8.505& 8.918 & 22.37 \\
  $\SIh$           &3.302 & 7.793& 7.853 & 19.50 \\ \hline
  \end{tabular}
   \label{tab-samp}
 \end{table}

We list the explicit values of masses and observables at the sample points
in Table~\ref{tab-samp}. 
We can see that the A-term is same order as other input parameters,
but the Higgs boson mass is about 125 GeV
owing to the suitable wino-to-gluino mass ratio.
The top squark mass is about 1.5 TeV
and the gluino mass is 2-3 TeV,
so that they could be in the reach of the HL-LHC. 
The bino and wino masses are between 2 TeV and 5 TeV
and they are far beyond the experimental reach of the LHC experiment.

When $|\mu|=250 (1000)$ GeV in the samples (a), (b), (c) and (d),
the thermal relic abundance is $\sim 0.01 (0.1)$. 
The self-annihilation rate
of the neutralinos in the zero-velocity limit,
denoted by $\langle \sigma v \rangle_0$, 
is $\mathcal{O}(0.1-1.0)\times 10^{-25} [{\rm cm}^3/s]$
and they are dominantly decaying to weak gauge bosons.
These processes are induced by the t-channel
neutralino or chargino exchange,
and then the rate is determined by the higgsino mass itself.  
These are important for the indirect detections as discussed below.

We also show the spin-dependent and spin-independent
LSP-proton cross sections, $\sigma_{\rm SD},\ \sigma_{\rm SI}$,
calculated by using micrOMEGA-4.2.5~\cite{Belanger:2013oya}.
$\sigma_{\rm SI}^h$ is obtained
from Eqs.~(\ref{eq-SIsimple}) and (\ref{eq-chxx}),
where $f_{T_q}^p$ are taken same as the values
adopted in micrOMEGA~\cite{Belanger:2014vza}.  
We can see the SI cross section is well described
by the tree-level Higgs-exchanging process,
but there are small deviations from the results of micrOMEGA.

A dominant source for the deviation come 
from the QCD corrections to the heavy quark matrix elements~\cite{Belanger:2008sj},
which enhance the cross section about 10$\%$ against the tree-level contribution.   
Besides, 
the top squarks could give contribution to the cross section,
when a mass difference 
$m^2_{\tilde{t}_1}-m^2_{\tilde{\chi}}$ is small. 
However, it is known that the leading contribution,
which is suppressed by $(m^2_{\tilde{t}_1}-m^2_{\tilde{\chi}}) m_t^2$, 
is proportional to the size of non-trivial mixing of the top squarks~\cite{Drees:1993bu}. 
The top squark is almost right-handed in our scenario 
and thus such contribution can not be sizable. 
We take the top squark corrections derived in Ref.~\cite{Drees:1993bu} into account, 
and confirm that these are about 1$\%$ against the tree-level countribution
at the sample~(d) and fewer for the other sample points.  
We have checked that
our results agree with the results of micrOMEGA exhibited in Table~\ref{tab-samp} 
within several $\%$-level after including these effects.  
There are potentially sizable corrections
from neutralino/Z-boson and chargino/W-boson mediated loop diagrams, 
where the neutralino and chargino are higgsino-like, 
but these are almost canceled out among them as shown in Ref.~\cite{Hisano:2011cs}.

\subsection{Indirect detection} 
Let us comment on indirect detections for the dark matter.
A pair of neutralinos decay to $W^+ W^-$ or $ZZ$
with the zero-velocity cross section: that is $\mathcal{O}(10^{-25})[{\rm cm}^3/s]$ as shown in Table~\ref{tab-samp}.

One of the most promising observables
may be the neutrino flux from the sun.
The capture rate of neutralino by the sun
is determined by the interaction
between neutralino and nucleons.
Since the spin-dependent cross section
is much larger than the spin-independent one,
the observations would give significant bounds on the spin-dependent cross section. 
The weak bosons produced by the annihilation of dark matter decay to neutrinos. 
The observed limit of neutrinos given by the IceCube is 
$ 3.76 \times 10^{-5}$ pb when the dark matter mass is 500 GeV 
and they decay to W-bosons exclusively~\cite{Aartsen:2016zhm}.
This limit is comparable to the expected limit at the XENON1T~\cite{Garny:2012it}. 
We will see that exclusion limits for the parameter space
from the XENON1T are much weaker than limits from the spin-independent cross section,
so that the current limit from IceCube experiment can not be important one.

 \begin{figure}[!t]
 \centering
 \includegraphics[width=0.65\linewidth]{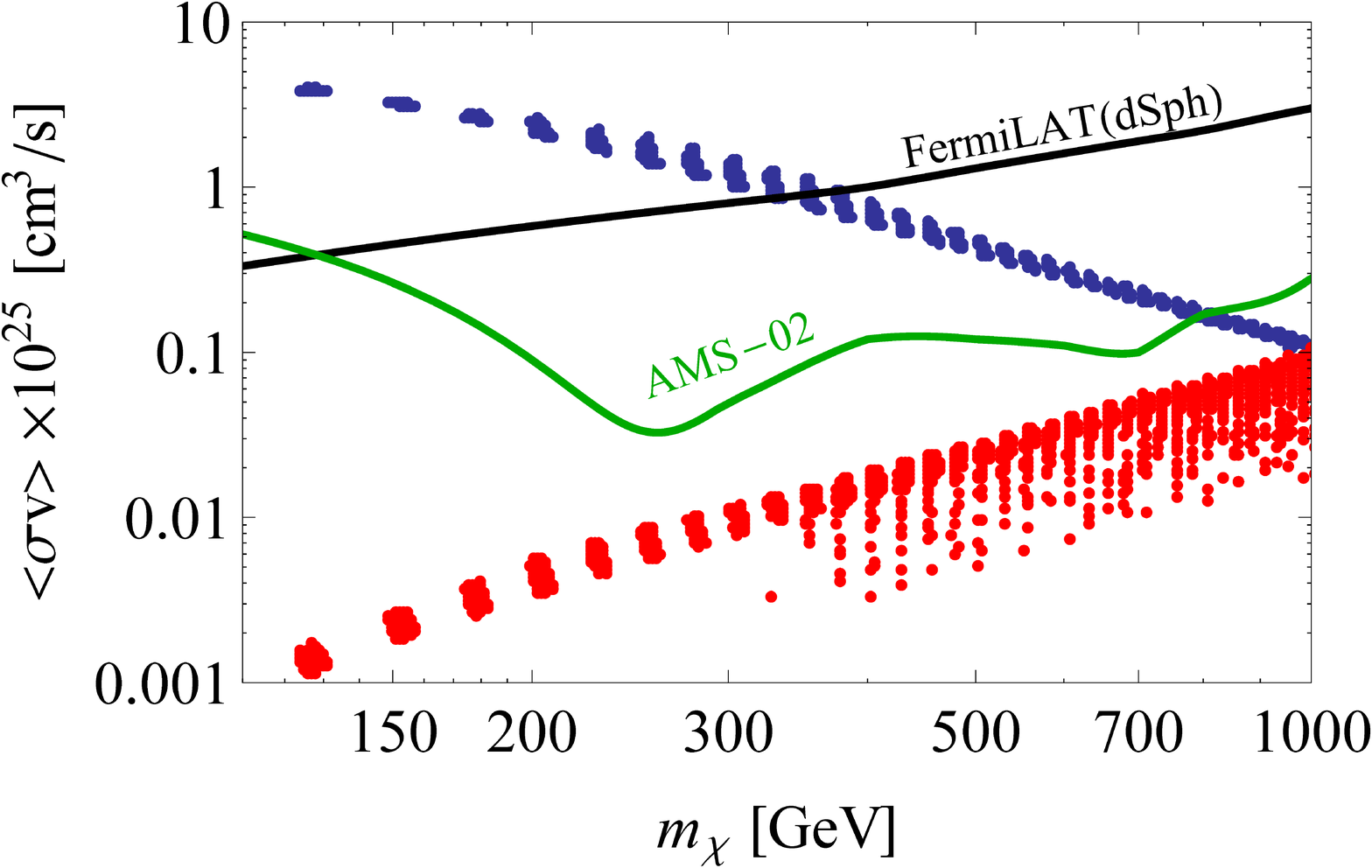}
 \caption{Exclusion limits and expected values in the NUGM scenario
  of the dark matter annihilation cross section.
  The blue (red) dots correspond to the non-thermal (thermal) scenario. }
 \label{fig-indirect} 
 \end{figure}

Cosmic ray observations such as photons, positrons and anti-protons
could be powerful tools to detect dark matter.
These limits of the annihilation cross section of DM
reach to $\mathcal{O}(10^{-25})[{\rm cm}^3/s]$
and the parameter region discussed in present paper is competing with these bounds.
We consider the recent experimental results
obtained by the Fermi-LAT~\cite{Ackermann:2015zua}
and AMS-02~\cite{Aguilar:2016kjl}. 
The former observes gamma rays
coming from the dwarf spheroidal satellite galaxies (dSphs) of the Milky Way
and the latter observes anti-protons
coming from dark matter annihilations in the Milky Way.
We refer the exclusion limit from the AMS-02 experiment
obtained in the analysis~\cite{Cuoco:2016eej}~\footnote{
Similar analysis is done in Ref.~\cite{Cui:2016ppb}.}. 
The Fermi-LAT experiment also observes gamma-rays
coming from the galactic center
and this potentially gives significant constraints
on the dark matter annihilation rate.   
However, the results are highly dependent
on dark matter density profiles~\cite{Gomez-Vargas:2013bea},
so that we do not discuss about this in present paper.

Figure~\ref{fig-indirect} shows
the upper limits on the annihilation cross section from the recent results of
the Fermi-LAT (black line) and the AMS-02 (green line). 
The dots are predictions from the NUGM scenario
and obtained by the parameter scanning to draw figures in next section.
We plot the points with $M_1 \ge 2.5$ TeV at the unification scale.
The blue dots indicate the lightest neutralino mass and the annihilation rate itself,
but it is multiplied by $(\Omega_{\rm \chi}/\Omega_{\rm obs})^2$ for the red dots. 
Since the higgsino-like dark matter dominantly annihilate
to W-bosons or Z-bosons by the t-channel exchange of the higgsino-like chargino or neutralino, 
the annihilation rate is mostly determined by the higgsino mass itself
and almost independent of other parameters. 
We see that the Fermi-LAT result excludes the neutralino lighter than about 300 GeV
and the AMS-02 excludes the neutralino lighter than about 800 GeV
in the non-thermal scenario.
On the other hand, the indirect detections do not give limits on the thermal scenario,
because the annihilation rate is suppressed
by the factor $(\Omega_{\rm \chi}/\Omega_{\rm obs})^2$. 
Exclusion limits on the higgsino dark matter produced from some non-thermal processes 
at the Fermi-LAT and the future planned CTA experiments~\cite{Carr:2015hta}
have been discussed in Ref.~\cite{Aparicio:2016qqb}.

\section{Numerical results}

Based on the above discussion, we summarize the experimental bounds
and show the allowed region.
As mentioned in Section 3.2, 
our analysis of the relic density includes two possibilities: thermal scenario and non-thermal scenario.
We calculate only thermal relic density and exclude the region with $\Omega_\chi h^2 > \Omega_{\rm obs} h^2$.
The difference of two scenarios only appear in the bound from the direct detection of DM.
$\Omega_\chi h^2 < \Omega_{\rm obs} h^2$ is possible in the thermal scenario, so
that the bound is relaxed.

 \begin{figure}[!t]
 \centering
 \includegraphics[width=0.98\linewidth]{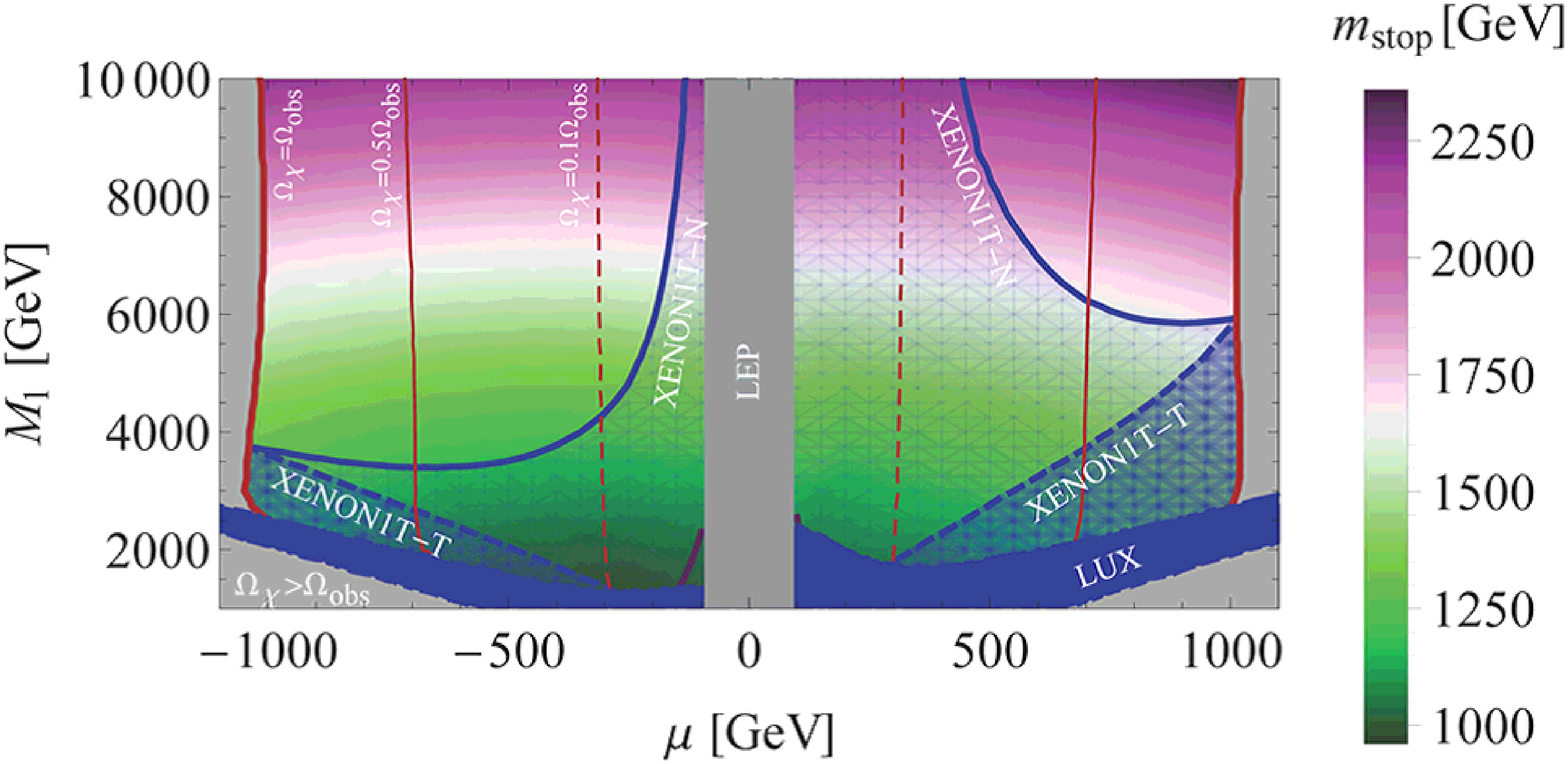}
  \caption{Values of the dark matter observables with $M_3=1.5$ TeV.}
 \label{fig-M1mu_M31500}
 \centering
 \includegraphics[width=0.98\linewidth]{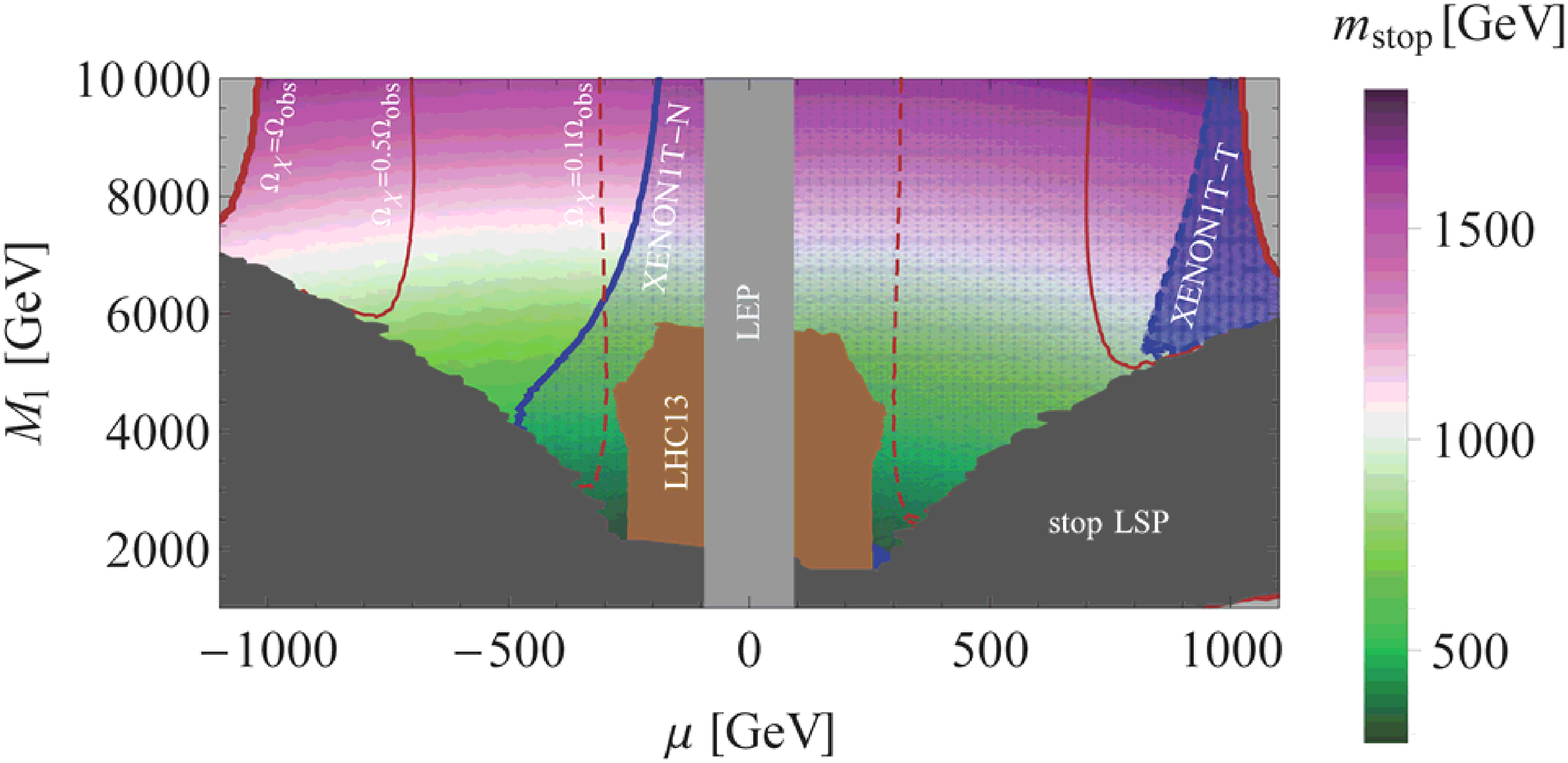}
  \caption{Values of the dark matter observables with $M_3=1.0$ TeV.} 
 \label{fig-M1mu_M31000}
 \end{figure}

 \begin{figure}[!t]
 \centering
 \includegraphics[width=0.98\linewidth]{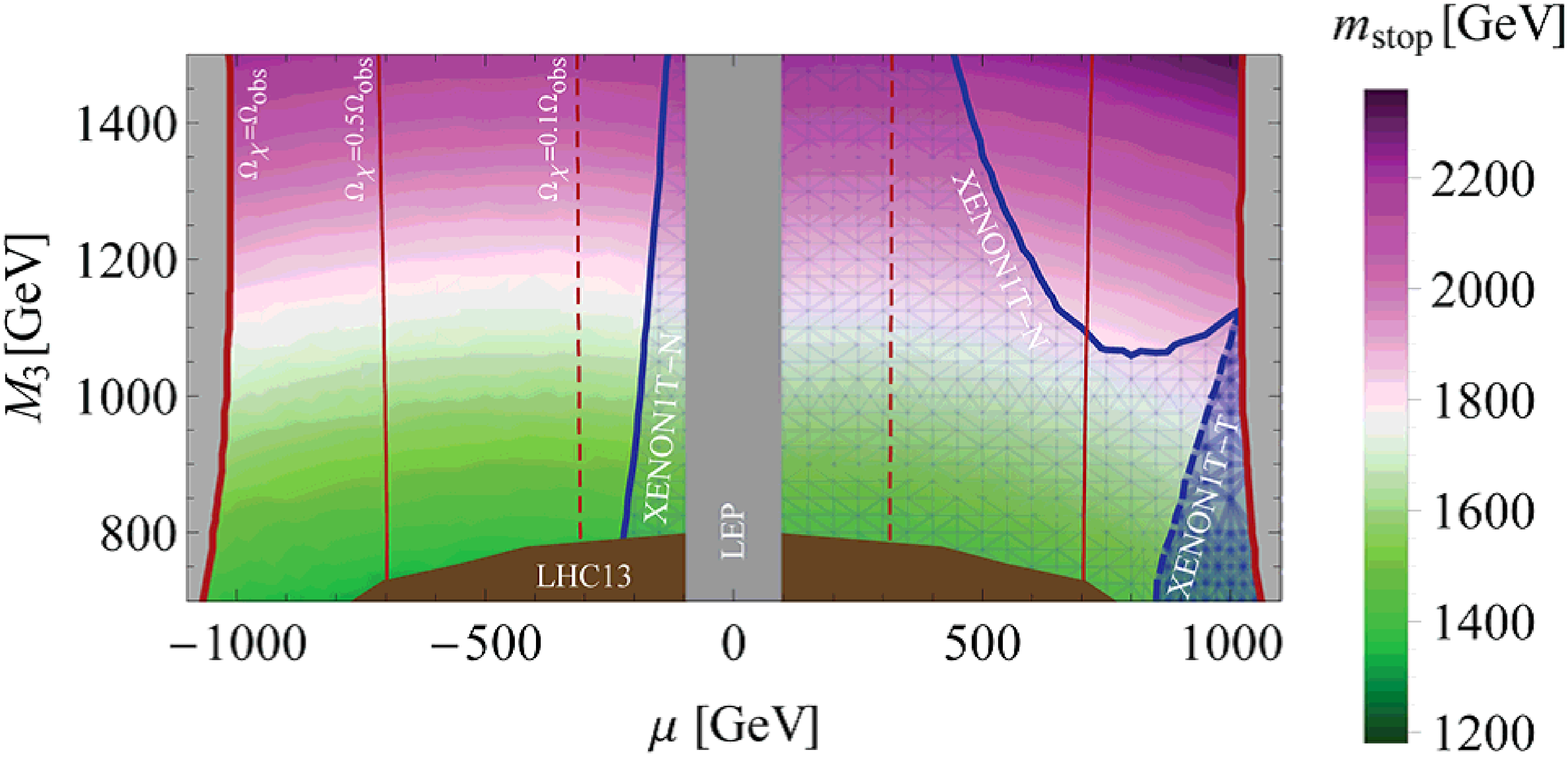}
  \caption{Values of the dark matter observables with $M_1=10$ TeV.} 
 \label{fig-M3mu_M110000}
 \centering
 \includegraphics[width=0.98\linewidth]{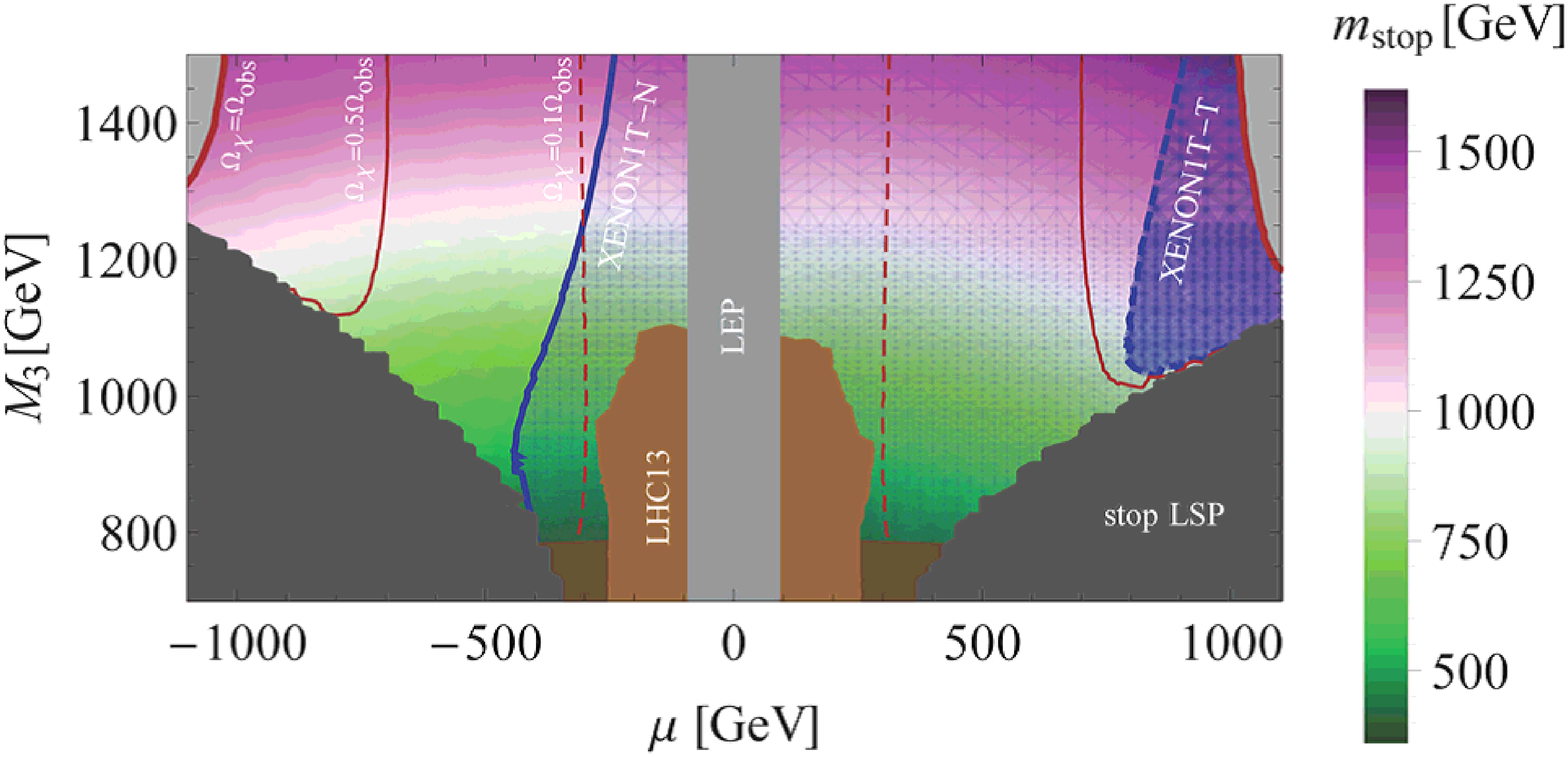}
  \caption{Values of the dark matter observables with $M_1=5.0$ TeV.} 
 \label{fig-M3mu_M15000}
 \end{figure}

 Figure~\ref{fig-M1mu_M31500} shows the allowed region for the dark matter observables,
 the top squark mass and exclusion limits from the collider experiments.
 We assume $m_0= 1$ TeV, $M_3=1.5$ TeV at the unification scale
 and $A_0, M_2$ are chosen to realize 
 the SM-like Higgs boson mass and the $\mu$-parameter at each point. 
 We take the ratio of the Higgs VEVs as $\tan\beta=10$.  
 We use softsusy-3.5.1~\cite{Allanach:2001kg} to calculate the RG effects
 and the mass spectrum of sparticles and Higgs bosons. 
 Their width and branching ratios are calculated
 by SDECAY and HDECAY~\cite{Djouadi:2006bz}. 
 The dark matter observables are calculated by micrOmega-4.2.5~\cite{Belanger:2013oya}.

 The red lines represent the thermal relic density of the neutralino,
 where the solid (dashed) lines correspond to
 $\Omega_\chi/\Omega_{\rm obs}=0.5\ (0.1)$ respectively.
 $\Omega_\chi = \Omega_{\rm obs} = 0.1188\pm0.0001$ \cite{Planck}
 is achieved in the red band around $|\mu| \simeq 1$ TeV. 
 The thermal relic density of the dark matter exceeds the observed value,
 $\Omega_\chi > \Omega_{\rm obs}$, in the light gray region,  
 so that this region is excluded
 if there is no dilution effects after the freeze-out of the neutralino. 
 The gray region at $|\mu| \le 90 $ GeV 
 is excluded by the LEP experiment~\cite{Jakobs:2001yg}.
 Although the charged and neutral components of higgsino are certainly degenerate,
 they can be probed by the mono-photon channel.
 The background color represent the mass of the lightest top squark. 
 The purple line around $M_1 \lesssim 2.0$ TeV and $\mu\simeq -100$ GeV
 is the expected exclusion limits for the spin-dependent cross section
 from the XENON1T experiment~\cite{Garny:2012it}.

 The spin-independent cross section exceeds
 the current limit given by the LUX experiment~\cite{LUX2016} in the blue band.
 This should be understood as the limits for the non-thermal scenario
 and the limit would be relaxed
 as the $\mu$-parameter decreases in the thermal scenario. 
 Such a suppression is, however, not so significant in this region,
 because the thermal relic density is enhanced
 due to the sizable fraction of bino to the lightest neutralino. 
 The blue shaded region covered by the solid blue lines (XENON1T-N) is the expected limit
 from the XENON1T experiment
 in the non-thermal case $\Omega_\chi = \Omega_{\rm obs}$,
 while the dashed blue line (XENON1T-T) corresponds to the same limit in the thermal case, 
 where the detection rate is suppressed due to the fewer neutralino relic density. 
 Note that the cross section of the spin-independent direct detection is always larger than $0.25\times 10^{-10}$ pb
 in all figures in this paper. Then, we expect that the future experiments, the XENON1T \cite{XENON1T} and the LZ~\cite{Akerib:2015cja}, 
 could cover our parameter region in the non-thermal scenario. 
 On the other hand, the current limit from the spin-dependent cross section
 is fully covered by the spin-independent one.

 The exclusion limit from the spin-independent cross section
 becomes stronger as the $\mu$-parameter decreases in the non-thermal scenario. 
 The reason is that the experimental limits for the cross section
 becomes tighter for lighter dark matter masses
 as long as the dark matter mass is heavier than about 40 GeV. 
 On the other hand, this effect is erased by the smaller LSP density
 $\Omega_\chi$ in the thermal scenario. 
 The light bino mass region is easier to be excluded
 due to the large bino-higgsino mixing,
 especially the well-tempered region has already excluded
 by the current LUX limit as well known.  
 The spin-independent cross section is significantly
 large for the positive $\mu$-parameter
 compared with the case of the negative $\mu$-parameter.
 This is because the cross section is proportional to
 $(1+{\rm sign}(\mu)\sin 2\beta)^2$ as can be read from Eq.~(\ref{eq-chxx}).

 Note that the exclusion limits on the $\mu$-$M_1$ plane are severer
 than the ones derived in Ref.~\cite{Cheung:2012qy}. 
 The difference comes from the fact that wino does not decouple completely in the NUGM scenario. 
 In order to keep the $\mu$-parameter smaller than 1 TeV,
 the wino mass at the unification scale has to be 3-4 times larger
 than the gluino mass. 
 The higher wino-to-gluino ratio is required for the lower typical sparticle scale
 which is defined as the geometric mean of the top squark masses.   
 In this case, ($M_2$, $M_3$) are about (4 TeV, 1.5 TeV) at the unification scale 
 and it enhances the spin-independent cross section.

 Figure~\ref{fig-M1mu_M31000} shows the allowed region for $\mu$ and $M_1$ at $M_3=1.0$ TeV.  
 The different value of $M_3$ influences to
 the direct detection rate and the top squark mass. 
 Top squark becomes the lightest SUSY particle in the dark gray region,
 and the top squark search at the LHC excludes the brown region. 
 The LHC bounds are projected from the analysis in Ref.~\cite{JK_PhDthesis}. 
 The bino mass has to be so large
 that top squark mass is larger than the higgsino mass.

 The lighter gluino mass leads the lighter wino mass  
 and the spin-independent cross section is enhanced
 by the wino-higgsino mixing.
 We see that the XENON1T experiment covers the whole region with $\mu > 0$
 in the non-thermal scenario.

 Figures~\ref{fig-M3mu_M110000} and \ref{fig-M3mu_M15000}
 show the allowed region for $\mu$ and $M_3$ 
 where $M_1$ is $5.0$ TeV and $10.0$ TeV at the unification scale, respectively.
 Other parameters are set to be the same as in Figures~\ref{fig-M1mu_M31500} and \ref{fig-M1mu_M31000}. 
 The constraint from the gluino search at the LHC is also applied to these figures
 and it excludes the dark brown region. 
 The gluino mass lower bound is around 800 GeV,
 so that there was no exclusion bounds in Figs.~\ref{fig-M1mu_M31500} and \ref{fig-M1mu_M31000}.
 We can see that experimental reaches from direct detections for the gluino mass
 can be much severer than those from the LHC experiment in the non-thermal scenario.

 The wino-higgsino mixing is reduced as gluino becomes heavy.
 The mixing, however, is not vanishing in our model-dependent analysis.  
 We see that the gaugino-higgsino mixing predicts
 the spin-independent cross section larger than $2.5\times 10^{-11}$ pb
 everywhere in all of the four figures.
 Thus the parameter region is on the neutrino floor~\cite{Billard:2013qya}  
 and the region in our analysis would be fully covered
 by the future planned observations such as the XENON-nT, LZD, PandaX-4T and so on.

\section{Conclusion}
In this paper, we study the dark matter physics
in the Non-Universal Gaugino Mass scenario.
The NUGM scenario is one of the possible setups of the MSSM  
to achieve  the 125 GeV Higgs boson mass and the $\mu$-parameter below 1 TeV, 
that naturally explain the origin of the EW scale. 
Since one top squark is relatively light in our scenario,
the authors in Refs. \cite{Abe:2015xva,Kawamura:2016drh}
study the current status and the future prospect on the direct search for top squark and gluino at the LHC. 

Although the higgsino mass is the most important
from the naturalness point of view,
higgsino can not be probed by the LHC
due to their suitable mass difference $\sim 2$ GeV.   
On the other hand,
the higgsino mass is critically important for dark matter physics
and can be tested by the dark matter observations.  
The higgsino mass can not be larger than 1 TeV
in order not to overclose the universe
if we assume that there is no dilution effect after the LSP is frozen out.

Direct detections for dark matter
are powerful tool to probe the neutralino sector of the MSSM. 
Even the bino and the wino masses are 3-4 TeV,
the spin-independent cross section
between higgsino and nucleon 
is in the observational reach.  
Therefore,
the wider parameter space can be covered by the direct detection
than the gluino search at the LHC,
when the wino-to-gluino mass ratio is fixed to realize the small $\mu$-parameter 
and the higgsino-like LSP dominates the relic density of dark matter.

If the neutralino density is determined
by the standard thermal process,  
the direct detection is sensitive to the parameter region where the higgsino mass is around 1 TeV, 
while the top squark and the gluino searches at the LHC
are generally sensitive to lighter higgsino.  
Thus the direct detection complement the direct search at the LHC.

The universal gaugino masses
are clearly disfavored by the recent dark matter observations.
The LSP is either bino or higgsino in this case,
but the bino LSP easily overclose the universe. 
Even if the higgsino LSP is realized
in some ways such as considered in Refs.~\cite{Feng:2012jfa,Baer:2012up}, 
light bino and wino are severely constrained
by the direct detections. 
The direct detection constraints push up
the gluino mass far above the experimental reach 
and such a heavy gluino indicates all other sparticles
are also hopeless to be discovered except in some special cases. 
Thus the non-universal gaugino masses
with relatively heavy bino and wino masses
seems to be more interesting than the universal gaugino masses.


\subsection*{Acknowledgments}
The work of J. K. was supported by Grant-in-Aid for
Research Fellow of Japan Society for the Promotion of
Science No. 16J04215.
%

\end{document}